\documentclass[prd,showpacs,letterpaper,nofootinbib,superscriptaddress,twocolumn]{revtex4}
 \usepackage{graphics}
 \usepackage{epstopdf}
 \usepackage{epsfig}
 \usepackage{amsmath,amssymb}

\begin{document}
\input{epsf}

\title{Using Atomic Clocks to Detect Gravitational Waves} 

\author{Abraham Loeb}

\affiliation{Astronomy Department, Harvard University, 60
  Garden St., Cambridge, MA 02138, USA}

\affiliation{School of Physics and Astronomy, Tel Aviv University, Tel
  Aviv 69978, Israel}

\author{Dan Maoz}

\affiliation{School of Physics and Astronomy, Tel Aviv University, Tel
  Aviv 69978, Israel}

\begin{abstract} 
Atomic clocks have recently reached a fractional timing precision of
$\lesssim 10^{-18}$. We point out that an array of atomic clocks,
distributed along the Earth's orbit around the Sun, will have the
sensitivity needed to detect the time dilation effect of mHz
gravitational waves (GWs), such as those emitted by supermassive black
hole binaries at cosmological distances. Simultaneous measurement of
clock-rates at different phases of a passing GW provides an attractive
alternative to the interferometric detection of temporal variations in
distance between test masses separated by less than a GW wavelength,
currently envisioned for the {\it eLISA} mission.

\end{abstract}

\pacs{04.80.Nn,95.55.Ym,95.85.Sz}
\date{\today}
\maketitle

\paragraph*{Introduction.} Over the past year, the precision of optical lattice 
clocks has advanced dramatically, to a fractional timing precision of
$(\Delta t/t) \sim 10^{-18}$, with prospects for a future improvement
by two additional orders of magnitude through the use of other atoms
\cite{Hinkley,Bloom,Beloy}. Here, we point out that the new regime of
timing precision made accessible by atomic clocks overlaps with the
expected amplitudes of time dilation and compression due to the
passage of gravitational waves (GWs). The standard time-dilation
effect for a clock at some distance from a black hole, would be
modulated by the periodic change in this distance due to the orbital
motion in a binary black hole system. Quantitatively, compact binaries
of supermassive black holes \cite{Sesana,Holz} at cosmological
distances produce a periodic modulation of the time-time component of
the metric (in a Newtonian gauge) at a level of
\begin{equation}
h_{00}\approx 9\times 10^{-18}
\left({D_L\over {\rm Gpc}}\right)^{-1} \left({{\cal M}_z\over
  10^6M_\odot}\right)^{5/3}\left({f\over {\rm mHz}}\right)^{2/3},
\label{eq:1}
\end{equation}
where $D_L(z)$ is the luminosity distance for a source at redshift
$z$, ${\cal M}_z\equiv (1+z) (M_1M_2)^{3/5}/(M_1+M_2)^{1/5}$ with
$M_1$ and $M_2$ being the masses of the binary members which are
assumed here to be on a circular orbit, and $f=(\omega/2\pi)$ is the
observed (redshifted) GW frequency (obtaining its maximum value based
on the orbital frequency at the binary's innermost stable circular
orbit).\footnote{In this paper, we adopt for pedagogical reasons a
  Newtonian gauge which is commonly used to describe the time-dilation
  effect due to stationary gravity, as measured in the Pound-Rebka
  experiment \cite{Pound}. In this gauge, an oscillating perturbation
  in the time-time component of the metric, $h_{00}$, would trigger
  periodic variation in the Pound-Rebka time dilation and a mismatch
  between the ticking rate of clocks separated apart. } Such black
hole binaries are a natural consequence of galaxy mergers
\cite{Colpi}. Around the Milky Way's supermassive black hole, GWs with
a similar amplitude and frequency can be emitted by orbiting close-in
stars \cite{Freitag}.  Such sources constitute the primary targets for
the future {\it eLISA} space observatory \cite{eLISA}, which aims to
detect GWs in the frequency range of 0.1-100 mHz and is planned for
launch in two decades. {\it eLISA} is designed as a laser
interferometer that will record the phase shift introduced by a
passing GW as the wave induces a change in the spacetime curvature,
and hence a change in light travel time between its three test masses,
which are separated from each other by $10^6$~km (less than the GW
wavelength).  Here, we consider the alternative approach of detecting
the differential time dilation experienced by clocks located at
different phases of a passing GW. For this purpose, we propose using
an array of atomic clocks.

\paragraph*{Atomic clocks in space.} We envision a set of small orbiting units
(at a minimum two of them) equipped with atomic clocks, and a primary
spacecraft between them, as illustrated in Fig. 1. For simplicity, the
clocks are assumed to be distributed along the circular orbit of the
Earth around the Sun at an orbital radius of 8.3 light-minutes
(1AU$\equiv 1.5\times 10^8$km), since this configuration minimizes the kinetic
energy requirements for launch. The typical distances between units
will thus be $\sim 10^8$~km. Additional units will naturally improve
the sensitivity and directional angular resolution.

\begin{figure}[th]
\includegraphics[scale=0.4]{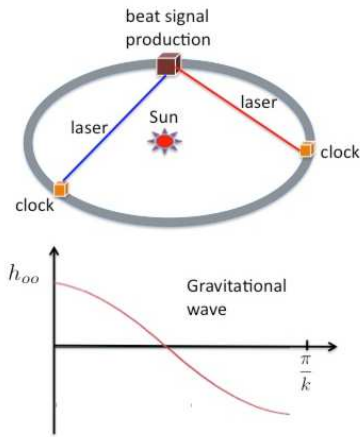}
\caption{Schematic illustration of an array of atomic clocks in a
  circular orbit around the Sun, as an observatory for low-frequency
  GWs. Clocks separated by half a wavelength, ${1\over
    2}\lambda=\pi/k= 1{\rm AU} (f/{\rm mHz})^{-1}$, of a passing GW,
  would show the largest difference in fractional timing, $h_{00}$,
  varying periodically over the GW period, $f^{-1}=2\pi/kc$. We adopt
  a Newtonian gauge for pedagogical reasons, but the measured physical
  effect should be independent of gauge.}
\label{Fig1}
\end{figure}

A passing GW characterized by a metric perturbation
$h_{\mu\nu}e^{i({\bf k}{\bf r}-\omega t)}$, will induce periodic
variations (over a period $2\pi/\omega=2\pi/kc$) in the ticking rate
of each clock \cite{Finn}. The amplitude of the timing variation would
be ${1\over 2}h_{00}$ (since time is dilated by
$({1+h_{00}})^{-1/2}$). For each pair of clocks, the relative phase of
the periodic variation $\Delta \phi={\bf k}\Delta {\bf r}$ will be
dictated by the projection of the difference in clock position vectors
$\Delta {\bf r}$ on the GW wavevector ${\bf k}$ (which defines the GW
propagation direction). The largest timing phase difference would be
realized between clocks separated by half a wavelength, ${1\over
  2}\lambda={\rm 1AU}(f/{\rm 1 mHz})^{-1}$, corresponding to a timing
phase difference $\Delta\phi=\pi$. Measurements of the $N(N-1)$ phase
differences $\left\{\Delta \phi_{i,j}\right\}_{i,j=(1,2,...N)}$ for a
set of $N$ clocks would allow to localize the GW source position on
the sky as $-{\bf k}$.

To communicate the periodic time dilation or compression signal
associated with a GW, the clock in each unit should drive an
optical-frequency ($\nu_{\rm laser}\sim 10^{15}$~Hz) laser, pointed at
the primary spacecraft. The required signal-to-noise ratio is not
dictated by the precision for distance measurements through laser
phase shifts, but rather by what is needed to maintain an optical
phase lock between the different units such that the ticking rate of
their clocks can be compared at a high precision. Techniques for
remote optical clock comparison had been developed over the past
decade \cite{Ye,Foreman} and are conceptually different from the
interferometric technique currently envisioned for eLISA.  In the
absence of phase noise in space, the distance noise level of $\sim
5\times 10^{-8}~{\rm cm/\sqrt{\rm Hz}}$ expected from the shot noise
of a 1 Watt laser detected by telescopes of 30 cm diameter across a
1-2AU path length, would be more than sufficient to phase lock a
single optical cycle. The needed laser is available and already in
use.  The primary spacecraft will interfere the laser signals from the
two daughter clocks to produce a beat signal with a frequency
$\nu_{\rm beat}\sim h_{00}\nu_{\rm laser}$, i.e. with a period of
$\lesssim 10^3$~s, comparable to the period of the GW, and hence
detectable.


\paragraph*{Discussion} The proposed method is fundamentally different from the 
current interferometric methods which underline the design of {\it
  Advanced-LIGO} \cite{LIGO} and {\it eLISA}. In GW interferometry,
the GW wavelength is considerably longer than the interferometer arm
length, and at any given moment one is essentially measuring the
roughly-uniform space curvature in the region between the test masses
by means of the light-travel time. Half a GW period later, the change
in light-travel time reveals the passage of the GW.  In our
clock-timing method, the separation between clocks is of order half a
GW wavelength and one measures the simultaneous difference in clock
rates between the low-curvature and high-curvature phases of the GW
(which again flips after half a GW period).  The change in the light
travel time between the clocks due to the GW averages out over a GW
wavelength, and thus amounts to a higher order correction to the beat
frequency, which can be neglected. 

Our proposal is to measure timing variations rather than distance
variations. Our detection scheme is not concerned with laser phase
variations that are induced by distance variations between
free-floating units, but rather with the change in the rate at which
clocks are ticking relative to each other. The lasers are locked to
the clocks so as to keep their timing stability at a high level.

The precision achieved by optical lattice clocks scales inversely with
the square root of the integration time \cite{Beloy} and limits the
clock-timing method to low-frequency GWs with $f\lesssim 1$ mHz,
allowing integration for $\sim f^{-1}\gtrsim 10^3$s. This, in turn,
requires clocks at $\sim 1$~AU separations, as considered here.
Secular trends (due to slow orbital variations in the gravitational
effect of the Sun or planets, the energy loss of the Sun, or the phase
shift due to the solar wind) over the short period of the measurement
($\lesssim 1$ hour) can be separated from the periodic GW signal in
the frequency range of interest here. Furthermore, for GW signals
lasting weeks or months, the orbital motion will scan the ecliptic,
improving the source localization accuracy. Solar oscillations at mHz
frequency (p-mode and especially g-mode) could also be detected by the
proposed array.

As indicated by Eq. (\ref{eq:1}), the timing precision of atomic
clocks is sufficient to detect a GW from a supermassive black hole
binary at cosmological distances within one GW period.  The
signal-to-noise ratio (SNR) of GW detection will improve in proportion
to the square-root of the number of wave cycles being
observed. Therefore, to achieve SNR$=1$, the fractional timing
precision per cycle could be worse than the GW amplitude by the
square-root of the number of wave cycles observed.

So far we assumed a monochromatic GW signal and ignored the slow drift
in the GW frequency (so-called chirp) due to the inspiral of the
supermassive black hole binary. The lifetime of tight binaries is in
fact limited by the rate of GW energy loss \cite{Peters}. For a
circular binary orbit, the fractional frequency increase during an
infinitesimal observing time $\Delta t_{\rm obs}$ is given by
\cite{Sesana},
\begin{equation}
{\Delta f\over f}= 0.3 \left({{\cal M}_z\over
  10^6M_\odot}\right)^{5/3} \left({f\over {\rm mHz}}\right)^{8/3}
  \left({\Delta t_{\rm obs}\over 1~{\rm hour}}\right) .
\end{equation}
The temporal rise in frequency $f$ (chirp) of the GW signal can also
be detected through clock timing to provide a second constraint on the
values of ${\cal M}_z$ and $D_L$ in addition to
Eq. (\ref{eq:1}). Since there are $\gtrsim 10^{12}$ galactic mergers
within a Hubble time in the observable volume of the universe, the
duty cycle of detectable signals (event rate times lifetime) could be
high \cite{Colpi,Wyithe,Loeb}.

The use of a large number of uncorrelated clocks, $N$, in every clock
unit, can improve the timing precision by a factor of
$1/\sqrt{N}$. This may become another advantage of the proposed
method, as atomic clocks become progressively small, low-weight and
inexpensive.  If the clocks are quantum entangled, the timing
precision could improve as $1/N$ and inversely with integration time
\cite{Komar}.

\bigskip
\paragraph*{Acknowledgments.}
We thank Pete Bender, Bence Kocsis, Pawan Kumar, Misha Lukin, Ron
Walsworth, and Jun Ye for helpful comments on the manuscript.
A.L. acknowledges generous support from the Sackler Professorship by
Special Appointment at Tel Aviv University (TAU), the Raymond and
Beverly Sackler TAU-Harvard program in Astronomy, and the NSF grant
AST-1312034.


\begin{references}

\bibitem{Hinkley} N. Hinkley, J. A. Sherman, N. B. Phillips,
  M. Schioppo, N. D. Lemke, K. Beloy, M. Pizzocaro, C. W. Oates, \&
  A. D. Ludlow, Science {\bf 341}, 121 (2013).

\bibitem{Bloom} B. J. Bloom, T. L. Nicholson, J. R. Williams, S. L.,
  Campbell, M. Bishof, X. Zhang, W. Zhang, S. L. Bromley, \& J. Ye,
  Nature {\bf 506}, 71 (2014).

\bibitem{Beloy} K. Beloy, N. Hinkley, N. B. Phillips, J. A. Sherman,
  M. Schioppo, J. Lehman, A. Feldman, L. M. Hanssen, C. W. Oates, \&
  A. D. Ludlow, Phys. Rev. Lett. {\bf 113}, 260801 (2014).

\bibitem{Sesana} A. Sesana, Classical and Quantum Gravity {\bf 30},
  244009 (2014); arXiv: 1307.4086

\bibitem{Holz} D. E. Holz, \& S. A. Hughes, Astrophys. J. {\bf 629}, 15
  (2005).

\bibitem{Pound} Pound, R. V., \& Rebka, G. A., Phys. Rev. Lett. {\bf
  3}, 439 (1959).

\bibitem{Colpi} M. Colpi, \& M. Dotti, Advanced Sci. Lett. {\bf 4},
  181 (2011).

\bibitem{Freitag} M. Freitag, Astrophys. J. {\bf 583}, L21 (2003).

\bibitem{eLISA} https://www.elisascience.org

\bibitem{Finn} M. J. Koop, \& L. S. Finn, Phys. Rev. {\bf D 90},
  062002 (2014).

\bibitem{Ye} J. Ye, J. L. Peng, R. J. Jones, K. W. Holman, J. L. Hall,
  D. J. Jones, S. A. Diddams, J. Kitching, S. Bize, J. C. Bergquist,
  L. W. Hollberg, L. Robertsson, \& L.-S. Ma, J. Opt. Soc. Am. B {\bf
    20}, 1459 (2003).

\bibitem{Foreman} S. M. Foreman, K. W., Holman, D. D. Hudson,
  D. J. Jones, \& J. Ye, Rev. Sci. Instrum. {\bf 78}, 021101 (2007).


\bibitem{LIGO} https://www.advancedligo.mit.edu

\bibitem{Peters} P. C. Peters, Phys.  Rev. {\bf 136}, 1224 (1964).

\bibitem{Wyithe} J. S. B. Wyithe, \& A. Loeb, Astrophys. J. {\bf 590},
  691 (2003).

\bibitem{Loeb} A. Loeb, Phys. Rev. {\bf D81}, 047503 (2010).

\bibitem{Komar} P. K\'om\'ar, E. M. Kessler, M. Bishof, L. Jiang,
  A. S. Sorensen, J. Ye, \& M. D. Lukin, Nature Physics {\bf 10}, 582
  (2014).

\end{references}
\end{document}